\documentclass[showpacs,twocolumn,amssymb,prl,aps]{revtex4}
\usepackage{graphicx,epsfig}

\begin{document}
\title{Polymorphism of the glass former ethanol confined in mesoporous silicon}
\author{Anke Henschel}
\author{Klaus~Knorr}
\author{Patrick~Huber}
\email[E-mail: ]{p.huber@physik.uni-saarland.de}
\affiliation{Experimental Physics, Saarland University, D-66041 Saarbr\"ucken, Germany}

\begin{abstract}
X-ray diffraction patterns of ethanol confined in parallel-aligned channels of $\sim$ 10~nm diameter and 50~$\mu$m length in mesoporous silicon have been recorded as a function of filling fraction, temperature and for varying cooling and heating rates. A sorption isotherm, recorded in the liquid state, indicates a three monolayer thick, strongly adsorbed wall layer and a capillary condensed fraction of molecules in the pore center. Though the strongly adsorbed film remains in an amorphous state for the entire temperature range investigated, the capillary condensed molecules reproduce the polymorphism of bulk solid ethanol, that is the formation of either crystalline or glass-like states as a function of cooling rate. The critical rate necessary to achieve a vitrification in the mesopores is, however, at least two orders of magnitude smaller than in the bulk state. This finding can be traced both to pure geometrical constraints and quenched disorder effects, characteristic of confinement in mesoporous silicon.
\end{abstract}

\maketitle

\section{Introduction}
Crystallization of molecular assemblies plays a dominant role for the usage of mesoporous matrices as hard templates for the preparation of nanosopic structures, ranging from nanorods and nanowires to more complex structured hybrid materials \cite{Dickinson2006, Thomas2008}. Also from a more fundamental point of view, it is of interest which architectural \cite{Knorr2008} and thermodynamical principles \cite{AlbaSim2006, Christenson2001} of the bulk state survive upon solidification in extreme spatial confinement of mesoporous hosts. For example, it has been demonstrated in a seminal work by Jackson and McKenna \cite{Jackson1991}, that the glass transition can be shifted down markedly upon spatial confinement on the mesoscale - see \cite{Alcoutlabi2005}.
\begin{figure}[hbt]
\begin{center}
\includegraphics[scale=0.4]{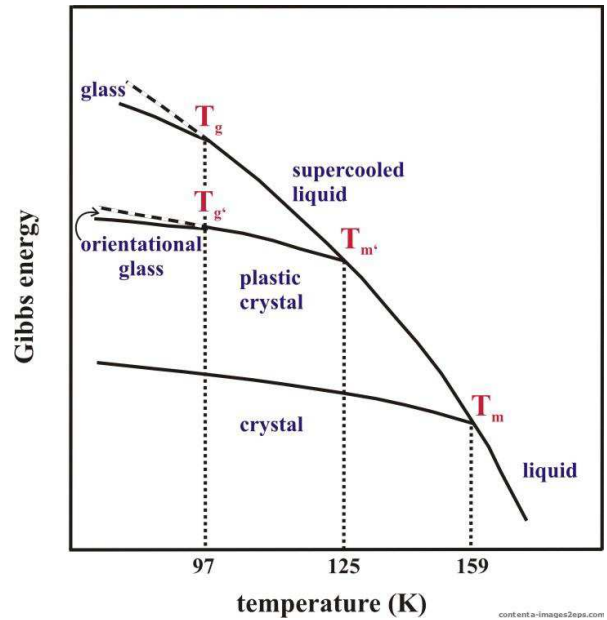}
\end{center}
\caption{\label{Phase} Phase diagram of bulk ethanol according to \cite{Viera}. Plotted are the Gibbs free energies of the distinct ethanol phases as a function of temperature and cooling rate (see discussion in the text). Note that, compared to the originally published version of the phase diagram \cite{Viera}, we exchanged the position of the ergodic (solid lines) and non-ergodic (dashed lines) of the glassy and orientational glass state according to the established thermodynamical models of the vitreous state \cite{Jaeckle1986}.}
\end{figure}

When molecular condensates solidify in mesopores, they often obey primary structural principles of the bulk solid state. Thus pore confined Ar \cite{Huber} is still cubic close packed (fcc) and the n-alkanes still form layered crystals \cite{Henschel}. Nevertheless there can be differences with respect to the bulk state. For N$_{\rm 2}$ embedded in a xerogel with a pore diameter of 7\,nm, the structural phase transition from the orientationally disordered hcp "plastic" high-temperature phase into the orientationally ordered Pa3 (cubic) low-temperature phase is suppressed \cite{Huber1998, Huber1}. Medium length neat alcohols, that are linear hydrocarbons with a terminating OH-group, still form layered crystals in mesopores \cite{Henschel1}. Albeit, a peculiar texture has been found, where the chains` long axes are oriented perpendicularly to the channel axis. Moreover, the phases with a final tilt of the molecule axis with regard to the layer normal, characteristic of the low-temperature state of these systems, are suppressed in mesopore confinement, in particular in mesoporous silicon \cite{Berwanger2009}.

In the present article we extend the previous studies on the phase transformation behaviour of medium-length alcohols in porous silicon towards the short-chain alcohol ethanol. Bulk solid ethanol is an interesting system that shows a peculiar type of polymorphism \cite{Haida, Viera, Ramos1997, Surovtsev2003, Ramos2006, Jiminez2007}, see Fig.\,\ref{Phase}. When cooled down in a sufficiently slow way, bulk ethanol crystallizes in a monoclinic structure. Here half of the molecules are in a trans-, the other half in a gauche-conformation. There is residual conformational, but no orientational entropy. On fast cooling (faster than a critical rate $r_{\rm c}$, $r_{\rm c}=6$\,K/min \cite{Ramos2006}), the liquid-monoclinic transition is bypassed, and ethanol finally forms a glass state, with frozen-in translational, orientational, and perhaps also conformational disorder ("structural glass"). For intermediate cooling rates and/or special annealing procedures, the liquid crystallizes in another modification, with a bcc center-of-mass lattice and disordered orientations of the molecules ("plastic phase"). At lower $T$, the orientations freeze-in, but the center-of-mass lattice remains bcc. This state is reminiscent of "orientational glasses" \cite{Hoechli1990}, known from so-called mixed crystals, and has been named "glassy crystals" \cite{Haida} or orientationally disordered crystals \cite{Ramos2006} for one component systems. Thus there are two liquid-to-crystal (at temperatures $T_{\rm m}$ and $T_{\rm m'}$) and two glass transitions at $T_{\rm g}$ and $T_{\rm g'}$ (see Fig.\,\ref{Phase}).

\section{Experimental}
The mesoporous substrate has been prepared by electrochemical etching of a p-doped (100) wafer, following a standard recipe \cite{Lehmann}. The pores are all parallel, perpendicular to the wafer surface with a length of approx. 50 $\mu$m determined by the etching time. The pore walls are relatively rough \cite{Kumar}. The mean pore diameter is 9.5($\pm$1.0)\,nm and the porosity 50\,\%, as estimated from a routine analysis of a N$_{\rm 2}$ sorption isotherm at 77\,K. The Si substrate has been mounted in a Cu cell, equipped with Be windows allowing the passage of the incoming and the diffracted x-ray beam. The cell is mounted to the cold plate of a closed cycle refrigerator. The maximum cooling rate $r$ in the $T$-range of interest - from 150\,K to 80\,K - is 0.5\,K/min. Moreover, the cell is connected to an all-metal gas handling setup, which allows filling and emptying the mesopores in a controlled manner through the vapour phase and thus to record in-situ ethanol sorption isotherms.
Highly purified ethanol was purchased from Riedel-de Haen with a quoted overall impurity concentration of less than 0.2\%. Since the vapour pressure of water at room temperature is approx. half the one of ethanol, the filling of the mesoporous matrix via the vapour phase further decreases the water content in the pore condensate. We estimate it to be significantly below 0.1\%. 
Cooling and heating scans have been interrupted and x-ray diffraction patterns have been recorded. Typically a pattern is recorded in $\Delta t=4$\,hours. In these cases the effective cooling and heating rates are much smaller, $r=\Delta T/\Delta t$, e.g. 0.04\,K/min for $\Delta T=10$\,K, $\Delta T$ is the temperature difference between consecutive diffraction patterns.

\begin{figure}[hbt]
\begin{center}
\includegraphics[scale=0.4]{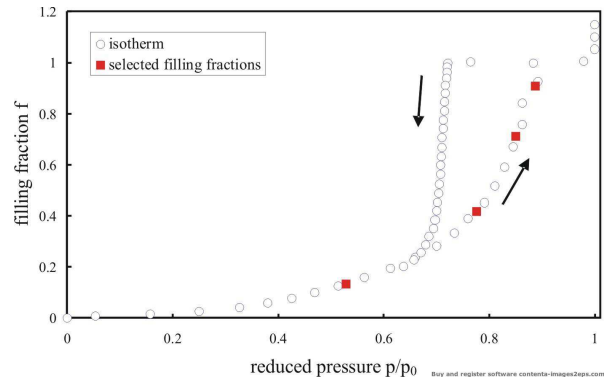}
\caption{\label{isotherm} Volumetric sorption isotherm of ethanol recorded at 277\,K. Indicated are the reduced vapour pressures versus the filling fraction of the mesopores. The hysteretic part is characteristic of the capillary condensed state, whereas the non-hysteretic part is typical of the film-condensed state as discussed in the text. X-ray diffraction patterns as a function of temperature were performed at the filling fractions marked by solid symbols.}
\end{center}
\end{figure}

An ethanol adsorption-desorption isotherm has been measured at 277\,K, i.e. in the liquid regime. Plotted in Fig.\,\ref{isotherm} is the filling fraction $f$ of the mesoporous matrix as a function of reduced vapour pressure $p/p_{\rm 0}$ upon filling and emptying of the pores, where $p_{\rm 0}$ refers to the bulk vapour pressure of ethanol of 13 mbar. Diffraction patterns on selected filling fractions ($f=$0.13, 0.42, 0.71, 0.91) have been recorded as a function of $T$ by means of coupled $\Theta$-2$\Theta$ scattering angle scans, which corresponds to wave vector transfers of $q=\frac {4 \pi} {\lambda} \sin(\Theta)$ with $\lambda=1.542$\,\r A.

\section{Results and discussion}
The measurement of a volumetric sorption isotherm is an important step for the characterization of the sample. Sample preparation by vapour adsorption leads to a homogeneous (coarse grained) distribution of the ethanol across the pore space \cite{Naumov2008}. The knowledge of the isotherm allows one to distinguish between the wall coating at lower $f$, here $f=0.13$, and the additional capillary condensate at higher $f$, here $f=$0.71 and 0.91, where the liquid fills parts of the pore centers, while formation of concave menisci. The $f$-value at which capillary condensate first appears on adsorption is, however, not exactly known, because of the fact that the adsorption branch of the adsorption-desorption loop is smeared out due to the finite width of the pore diameter distribution. In particular it is a priori not clear whether there is already a small fraction of capillary condensate present for $f=0.41$. Finally and perhaps most importantly in the present context, one can be sure for all $f$-values investigated that there is no bulk condensate outside the pores, the solidification of which could easily act as nucleus for the solidification of the pore filling.

\begin{figure}[hbt]
\begin{center}
\includegraphics[scale=0.4]{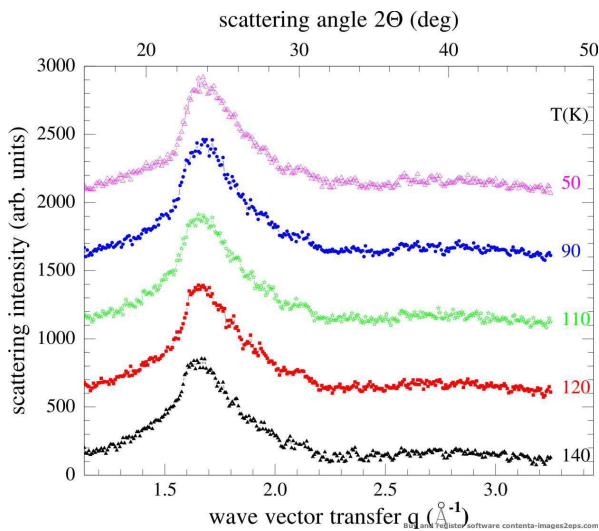}
\caption{\label{amorph} X-ray diffraction patterns of confined ethanol recorded for a filling fraction $f=$~0.91 upon cooling with 0.08\,K/min. Plotted is the scattered intensity both as a function of the wave vector transfer (lower abscissa) and scattering angle (upper abscissa) for selected temperatures as indicated in the figure.}
\end{center}
\end{figure}

For all x-ray measurements the scattering vector $q$ was oriented perpendicular to the wafer surface and hence parallel to the pore axis. Here the diffracted intensity of the empty sample ($f=0$) served as a background and has been subtracted. The diffraction patterns for $f=$0.13 and 0.42 just show two broad maxima. The first one is centered at about $q=$1.65 $\r A^{-1}$. The second one extends from about 2.4$\r A^{-1}$ to 3.4$\r A^{-1}$ in $q$. Apart from a small shift of the first maximum, these features are independent of $T$. They are interpreted as the first and second maximum of the structure factor of a liquid or glass like state of the adsorbed film on the pore walls, the maximum thickness of which corresponds to about three monolayers. The patterns (not shown) are identical to the ones known of amorphous ethanol - see discussion below and Fig. \ref{amorph}. Note that even for Ar, one of the worst candidates for glass formation, the transition of such a film into a crystalline state is suppressed by the strong interaction of the admolecules with the rough substrate \cite{Huber, Wallacher,Hofmann2005}. 
 
Depending on the time-temperature history the measurements with $f=$0.71 and 0.91 show different diffraction patterns. The experiment on the highest filling fraction ($f=0.91$) started with a cooling run directly down to 50\,K with the highest possible rate ($r=0.5$\,K/min). The pattern obtained at this temperature is devoid of Bragg peaks, demonstrating that the crystallization of pore confined ethanol can be suppressed by cooling down with rates that are considerably lower than the critical rate $r_{\rm c}$ of the bulk system. Upon heating, we did not observe a transition of the canonical glass phase into the plastic crystal, as it has been observed in the bulk state \cite{Ramos2006}. 
Thereafter diffraction patterns were recorded during the cooling/heating cycle, thereby the rate $r$ is considerably reduced. The diffraction patterns obtained for $f=0.91$ with an intermediate rate of 0.08\,K/min are of the liquid/glassy type with no Bragg peaks (see Fig.\,\ref{amorph}). Obviously even this rate is still fast enough to bypass the nucleation of the bcc and of the monoclinic phase and to establish the structural glass at low $T$.

\begin{figure}[hbt]
\begin{center}
\includegraphics[scale=0.5]{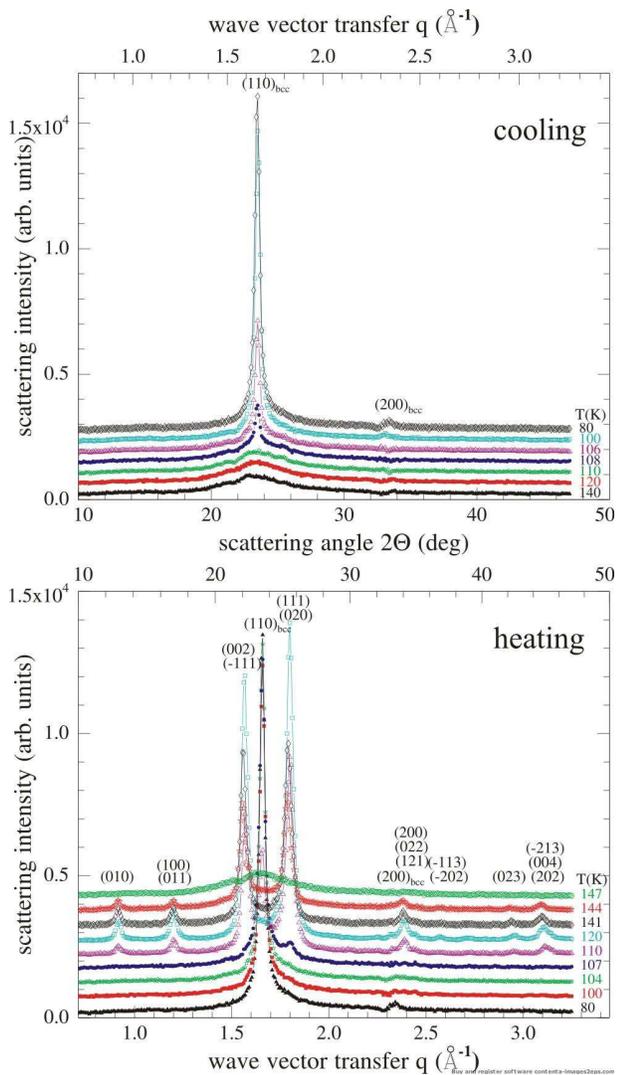}
\caption{\label{crystal} X-ray diffraction patterns of confined ethanol recorded at selected temperatures for a filling fraction $f=$~0.91 upon cooling (upper panel) and heating (lower panel) with 0.008\,K/min. Plotted is the scattered intensity both as a function of the wave vector transfer (lower abscissa) and scattering angle (upper abscissa) for selected temperatures as indicated in the figure.}
\end{center}
\end{figure}

Slow cooling-heating cycles from 160\,K down to 50\,K and back up to 160\,K have been performed with an effective cooling rate of 0.008\,K/min. The results on the two fillings $0.91$ and $0.71$ are identical apart from the higher diffracted intensity for $f=0.91$ and minor shifts of transition temperatures. In Fig.\,\ref{crystal} we show diffraction patterns for $f=0.91$. On cooling, the diffraction pattern of the liquid state is conserved down to 110\,K, well below the melting temperature $T_{\rm m}=159$\,K of the monoclinic phase of bulk ethanol. Around $108\pm1$\,K a strong Bragg peak appears that is easily identified as the leading (110) diffraction peak of the bcc phase \cite{Srinivasan, Fayos}. There is also a weak feature between 33\,deg and 34\,deg in $2\Theta$ which stems from the wings of the (200) reflection of the Si substrate and at least at lower $T$ from the (200) reflection of the bcc phase. There are no indications of further higher order bcc reflections. X-ray diffraction patterns on plastic crystal phases usually do not show more than one or two Bragg reflections. See ref.\,\cite{Srinivasan} for an x-ray diffraction pattern of the bcc phase of bulk ethanol. The rapid decay of the Bragg intensities with increasing scattering angle is due to the combined effect of the molecular form factor and the large Debye-Waller factors of these partially disordered phases. 

The intensity of the Bragg intensities for $f=0.71$ and 0.91 scales with $f-f_{\rm c}$, $f_{\rm c}=0.3\pm0.1$, rather than directly with $f$. Analogous observations have been made in other pore condensates, Ar in particular \cite{Huber, Wallacher, Hofmann2005}. It testifies that the film condensed, amorphous fraction of the condensate keeps its identity when additional capillary condensate is added on top at higher $f$, which forms crystalline structures.  

On heating the bcc (110) peak disappears at a temperature slightly higher, by about 3\,K, than at which it has appeared on cooling, but clearly lower than the melting temperature $T_{\rm m'}$ of bulk bcc ethanol ($T_{\rm m'}=125$\,K). Simultaneously a series of diffraction peaks grows which is easily identified as that of the monoclinic phase known from the bulk state. A powder pattern has been calculated from the structural data of \cite{Joensson}. The Bragg angles of the present experiment agree with bulk data. The Bragg intensities differ somewhat, indicating the pore confined monoclinic crystallites have some preferred crystallographic orientation with respect to the pore axis, but definitely the texture is not as extreme as previously  observed for n-hexane \cite{Henschel2}, medium length n-alkanes \cite{Henschel} and n-alcohols \cite{Henschel1}. Despite a possible texture a comparison of the diffraction pattern with the ones typical of the four known monoclinic phases of ethanol indicates that the pore-confined phase refers to what has been termed the monoclinic $\alpha$ phase of ethanol \cite{Ramos2006}. The monoclinic phase melts around $147\pm1,5$\,K, well below the bulk melting temperature $T_{\rm m}=159$\,K.

\begin{figure}[hbt]
\begin{center}
\includegraphics[scale=0.4]{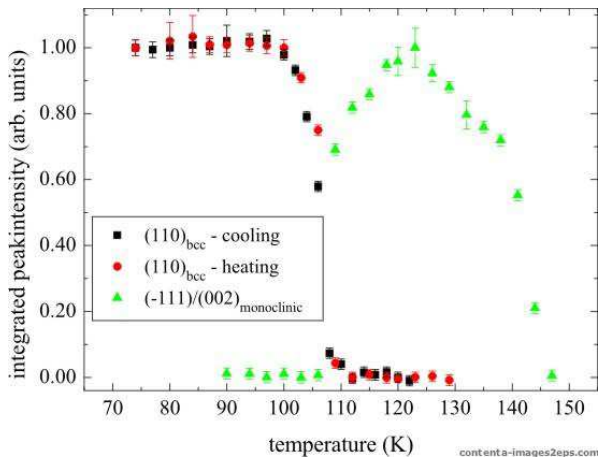}
\caption{\label{intensity} Bragg-intensities as a function of temperature of the (110) peak of the bcc phase (cooling and heating) and of (-111)/(002) peaks of the monoclinic crystal phase appearing on heating for a filling fraction $f=0.71$.}
\end{center}
\end{figure}

In Fig.\,\ref{intensity} the $T$-dependence of selected Bragg peaks for $f=0.71$ is plotted. The results supply values of three transformation temperatures: $T_{\rm 1}$, the melting temperature of the monoclinic phase, $T_{\rm 2}$ the solidification temperature of the bcc phase, and $T_{\rm 3}$ the temperature of the bcc-to-monoclinic transformation. $T_{\rm 1}$, $T_{\rm 2}$ and $T_{\rm 3}$ are about $147\pm1,5$\,K, $106\pm1$\,K, and $109\pm1,5$\,K, respectively. $T_{\rm 1}$ is 8\% lower than the melting temperature $T_{\rm m}$ of the monoclinic phase of bulk ethanol.  Similar reductions of the melting temperature upon pore confinement have been observed for other molecular condensates. For CO a reduction of 9\% has been found in the porous silica substrate SBA-15 with about the same pore diameter  \cite{Kityk}. 

The liquid-to-bcc transformation of the pore condensate occurs at a temperature $T_{\rm 2}$, 15\% below the reference temperature $T_{\rm m'}$ of the bulk system. For CO a reduction of the freezing temperature by 15\% with respect to the bulk state has been observed \cite{Wallacher1}. From the comparison with CO, one would expect a melting temperature of the pore confined bcc phase of about 116\,K (9\% above $T_{\rm 2}$). But the temperature $T_{\rm 3}$ is considerably lower. Furthermore the $T_{\rm 2}-T_{\rm 3}$ thermal hysteresis of about 2\,K is much smaller than any hysteresis that has been observed for first order phase transitions in mesopores. It appears that on heating the bcc phase transforms directly into the monoclinic phase and that this happens at a relatively low temperature. Obviously it is easier for the monoclinic groundstate to nucleate out of the bcc metastable state than out of the supercooled liquid. 

The two crystalline structures, bcc and monoclinic, observed for $f=0.71$ and 0.91, show wide single-phase-regions. Phase coexistence (bcc-liquid, monoclinic-liquid, bcc-monoclinic) is reserved for relatively narrow T-intervals, as is evident from the Bragg intensities in Figs. \ref{amorph} and \ref{crystal}. Obviously, a transformation process once started spreads out along the pores in spite of variations in size and shape of the pore cross section which can lead to blockades of the advancing transformation front \cite{Wallacher, Khokhlov}.

Note however that the transformation rate also depends on the characteristic time of the individual processes that contribute to the transformation that may be different in pore confinement. There has been a lot of work on pore confined glass formers, but a clear understanding has not evolved. In some systems the structural relaxations are faster, in others slower than in the bulk state \cite{Jackson1991, Park}. This leads to downward or upward shifts of the apparent glass transition temperature T$_{\rm g}$. Molecular Dynamics computer simulations on a binary van-der-Waals condensate show that the nature of the confining walls are of importance, smooth walls speed up the relaxation, rough walls slow it down \cite{Scheidler}.

The critical cooling rate that is required to bypass crystallization is much smaller in the bulk state since freezing of the ethanol pore liquid into the monoclinic phase was never reproduced for even the slowest cooling rate applied. Sufficiently fast cooling of ethanol established a structural glass. The critical cooling $r_{\rm c}$ of ethanol in porous silicon is between 0.008 and 0.08\,K/min, almost two orders of magnitude smaller than in the bulk state \cite{Viera}. To some part this stems from shifts of the free enthalpy $G$ and the chemical potential $\mu$ of the phases involved with respect to the bulk state. Upon pore condensation, $\mu$ is lowered, because of the attractive interaction with the substrate, the reduction being larger for the liquid than for the solid state, and larger for a plastic phase than for an orientationally ordered crystalline phase \cite{Huber1}. This is because the matching of the crystalline state to the pore geometry requires lattice defects, grain boundaries in particular. The energy costs for such defects are higher in crystalline as compared to a plastic phase. Thus the free enthalpy of the pore filling is modified with respect to the situation shown in Fig.\,\ref{Phase} in the sense that the differences $\Delta G$ between the competing phases liquid, bcc, monoclinic are reduced, the largest reduction occurring between the liquid and the monoclinic phase. Thereby the thermodynamic driving force for a transformation into a phase with lower $G$ is reduced. The reduction of the transition temperatures is a direct and well established consequence. Of course, reduced $\Delta G$'s should also lead to slower transformation kinetics.

Leaving aside these effects of pure geometrical constraints, it should also be kept in mind that the pore diameter variation within one tubular pore along with the sizeable roughness of the pore walls introduces a static variation in the interaction of a sizeable fraction of molecules with their neighborhood, which corresponds to random interaction fields \cite{Hoechli1990}. This type of quenched disorder has been found to affect first and second order phase transitions of pore condensates in mesoporous silicon significantly \cite{Wallacher1, Guegan2006, Naumov2008}. In general, it favors the disordered phases, in agreement with the present finding. In principal, substitutional disorder, that are impurities, can introduce similar random fields. In fact for ethanol with only a small water content of 0.5\%  a significantly increased vitrification tendency has been reported in the past \cite{Jiminez2007}. To the best of our knowledge, however, no such increased glass formation has been reported for ethanol with water impurities below 0.1\% so far, as investigated here. 

\section{Conclusions}
On the whole pore confined ethanol reproduces the polymorphism of the bulk state. The present diffraction study gives direct evidence of the two crystalline modifications, the monoclinic phase and the bcc plastic phase. The experiment cannot distinguish
between the liquid and the amorphous, structural glass phase, but it is plausible that the translational and orientational degrees of freedom freeze-in at low $T$ in a way similar to the bulk state, not necessarily at the transition temperatures $T_{\rm g}$ and $T_{\rm g'}$ of the bulk state. This quasi bulk behaviour is reserved to the condensate next to the central pore axes that is formed by capillary condensation. The adsorbate on the pore walls, which is the only condensate present at $f=0.13$ and maybe even 0.42, does never crystallize. The observations reported here may stimulate further investigations on confined short-length n-alcohols. In particular, it would be interesting how the phase transformation and vitrification behaviour changes towards longer alkyl-chains, where the polymorphism in the bulk state eventually vanishes and even at the highest experimentally achievable cooling rates no vitrification has been observed for bulk samples so far . Finally, the mesoscopic roughness and pore structure of mesoporous silicon can be tuned by relatively simple chemical and electrochemical means \cite{Lehmann, Kumar}, which may provide the possibility to study the interesting question, how the vitrification tendency of ethanol varies as a function of both magnitude of quenched disorder and mean pore diameter in the future.

This work has been supported by the Deutsche Forschungsgemeinschaft via the Sonderforschungsbereich 277, Saarbr\"ucken.

\end{document}